\documentclass[acmsmall]{acmart}

\usepackage{algorithm}
\usepackage{algpseudocode}
\usepackage{amsmath}
\usepackage{graphicx}
\usepackage{subcaption}
\usepackage{xspace}
\usepackage{multirow}
\usepackage[most]{tcolorbox}

\newboolean{showcomments}
\setboolean{showcomments}{true}         
\ifthenelse{\boolean{showcomments}}
  {\newcommand{\nb}[2]{
  \fbox{\bfseries\sffamily\scriptsize#1}
     {\sf\small$\blacktriangleright$\textit{\textcolor{red}{#2}}$\blacktriangleleft$}
   }
  }
  {\newcommand{\nb}[2]{}
   
  }

\newcommand\new[1]{{\color{black}#1}}

\AtBeginDocument{%
  }

\setcopyright{cc}
\setcctype{by}
\acmDOI{10.1145/3832208}
\acmYear{2026}
\acmJournal{PACMSE}
\acmVolume{3}
\acmNumber{ISSTA}
\acmArticle{ISSTA117}
\acmMonth{10}
\acmSubmissionID{issta26main-p1023-p}
\received{2026-01-30}
\received[accepted]{2026-04-16}

\begin{document}

\title{Testing Retrieval-Augmented Generation Systems with Chunk Coverage}


\author{Jinhan Kim}
\orcid{0000-0002-0140-7908}
\affiliation{%
  \institution{Università della Svizzera italiana}
  \department{Software Institute}
  \city{Lugano}
  \country{Switzerland}
}
\email{jinhan.kim@usi.ch}

\author{Samuele Pasini}
\orcid{0000-0002-7900-3727}
\affiliation{%
  \institution{Università della Svizzera italiana}
  \department{Software Institute}
  \city{Lugano}
  \country{Switzerland}
}
\email{samuele.pasini@usi.ch}

\author{Paolo Tonella}
\orcid{0000-0003-3088-0339}
\affiliation{%
  \institution{Università della Svizzera italiana}
  \department{Software Institute}
  \city{Lugano}
  \country{Switzerland}
}
\email{paolo.tonella@usi.ch}

\renewcommand{\shortauthors}{Kim et al.}


\begin{abstract}
Retrieval-Augmented Generation (RAG)-based systems\footnote{For brevity, RAG-based systems are referred to as RAG systems throughout this paper.} are increasingly deployed in high-stakes settings where correct behaviour depends not only on the language model but also on the retrieval component that selects external documents at inference time. While existing RAG evaluation metrics assess retrieval and generation quality on a per-query basis, typically relying on query-level test oracles such as reference answers or relevance annotations, they provide limited insight into whether a test suite adequately exercises the retrieval behaviour of the system as a whole. In this paper, we introduce Chunk Coverage (CC), an oracle-independent test adequacy criterion for testing the retrieval component of RAG systems. CC measures the fraction of corpus chunks that are retrieved at least once across a test suite, providing a structural view of which parts of the retrieval space have been exercised. We further show how CC can be used to guide test selection and generation by prioritising queries that expand coverage of previously unexercised retrieval regions. We evaluate CC on clinical and financial RAG system scenarios. CC-guided testing reaches 50\% of attainable coverage 1.7$\times$ faster than random selection and 4.2$\times$ faster than redundancy-biased strategies. Moreover, CC improves fault detection effectiveness (APFD) by 10\% to 25\% over random, indicating earlier discovery of distinct retrieval faults. These results show that CC captures retrieval diversity relevant to effective testing without requiring test oracles.
\end{abstract}



% \end{CCSXML}\ccsdesc[500]{Software and its engineering}



\maketitle

% !TEX root = ../main.tex

\section{Introduction}
\label{sec:intro}

Large Language Models (LLMs)~\cite{brown2020language} are increasingly deployed as components of complex software systems, particularly in domains where decisions are safety-critical, regulated, or high-stakes~\cite{liu2025scales, ni2025towards, gan2025retrieval}. In such settings, developers are not merely interested in average model performance, but in gaining confidence that the system behaves as intended across the situations it may encounter. As with any such system, systematic testing is therefore essential. At the same time, the increasing adoption of Retrieval-Augmented Generation (RAG)~\cite{lewis2020retrieval} as an architectural pattern in LLM systems introduces additional challenges for testing. RAG refers to a class of systems in which responses are generated based on information retrieved from an external document corpus at inference time. While RAG is most commonly instantiated in combination with LLMs, the retrieval component itself is conceptually independent of any particular generator.

In a typical RAG-based LLM system, test queries\footnote{Throughout the paper, we use the terms \textit{query}, \textit{test}, and \textit{test query} interchangeably to refer to the user-provided input that forms part of the prompt (e.g., within a prompt template) and triggers the RAG-based LLM system's response.} and corpus documents are embedded into a shared vector space, and a retriever selects a small set of documents that are incorporated into the prompt provided to the LLM. By grounding generation in an explicit document collection rather than relying solely on parametric knowledge acquired during training, such systems can operate over proprietary, dynamic, or domain-specific information sources, including internal technical documentation, scientific archives, legal texts, or organisational knowledge bases~\cite{ni2025towards}. In these deployments, correct system behaviour increasingly depends on which documents are retrieved, rather than on the LLM alone.

This architectural separation fundamentally changes what it means to test an RAG-based LLM system. Under a fixed configuration, retrieval behaviour is determined by the embedding model that maps queries and corpus chunks into a shared vector space, together with the vector database and its retrieval policy. These components jointly determine which parts of the corpus are accessible to the system at inference time. If relevant information is not retrieved, the LLM cannot compensate, regardless of its generative capabilities. Testing an RAG-based LLM system, therefore, necessarily entails testing its retrieval behaviour, not just the quality of generated outputs.

Despite this, the testing of RAG systems is still predominantly framed as an evaluation problem. Existing RAG evaluation metrics assess properties such as retrieval relevance, answer faithfulness, and grounding for individual query--response pairs, often relying on an auxiliary LLM as a proxy judge~\cite{es2024ragas, saadfalcon2023ares}. These metrics are valuable for benchmarking and comparing configurations, but they answer a different question: \textit{how well does the system perform on a given query?} They do not address a testing-native concern: \textit{does a test suite adequately exercise the retrieval behaviour of the system as a whole?} From a software testing perspective, this distinction is critical. A test suite may achieve strong average evaluation scores while repeatedly retrieving the same small subset of popular documents, leaving large portions of the corpus effectively untested. Such blind spots are problematic when validating retrieval configurations prior to deployment, particularly in high-stakes settings where previously unseen retrieval behaviour may surface only after release.

This gap arises because current evaluation practices focus on per-query outcomes, while retrieval behaviour is inherently a \textit{suite-level} phenomenon. Under a fixed configuration, the retriever induces a behavioural space over the document corpus: some regions are frequently retrieved, others rarely, and some not at all. Neither existing RAG evaluation metrics nor prior testing approaches make this retrieval-induced behaviour observable at the level of a test suite. What is missing is a test adequacy criterion that characterises how thoroughly the retrieval component has been exercised, independently of output correctness.

In this paper, we introduce \textit{Chunk Coverage (CC)}, a structural adequacy criterion for testing the retrieval component of RAG systems. CC measures the fraction of corpus chunks that are retrieved at least once across all tests in a suite. Analogous to code coverage in traditional software testing, CC abstracts away from output correctness and individual test outcomes. It is not intended to detect faults or replace per-query evaluation metrics. Instead, it provides a diagnostic view of what a test suite has exercised in the retriever. Importantly, CC is oracle-independent: it requires neither ground-truth relevance annotations nor reference answers, and can be computed solely from retrieval traces.

Coverage becomes practically useful when it can guide testing. We show how CC enables coverage-guided test selection and generation: as tests are executed, retrieved chunks are tracked, uncovered regions of the retrieval space are identified, and subsequent tests are prioritised to expand coverage. This process can be instantiated either by synthesising new queries using an auxiliary LLM or, when an existing query pool is available, by simulating test selection under different strategies. In both cases, CC provides the guidance signal while remaining independent of the system under test and of LLM-based judging.

We evaluate CC in two representative, retrieval-essential scenarios: clinical decision making over patient records derived from MIMIC-IV~\cite{hager_mimic-iv-ext_nodate, johnson2023mimic}, and financial question answering over heterogeneous documents from T2-RAGBench~\cite{strich2025t}. Across five datasets from both scenarios, coverage-guided generation consistently accelerates exploration of the retrieval space. Compared to random selection, CC reaches 50\% of the attainable coverage 1.7$\times$ faster, and 4.2$\times$ faster than deliberately overlap-biased generation.
% with an average speedup of 1.7$\times$ over random baselines.
% \paolo{I do not understand the difference between the $1.9\times$ speedup w.r.t. random selection and the $1.7\times$ speedup w.r.t. random baselines: it sounds like they are both random.}

Higher coverage is not merely a structural property. Using fault detection effectiveness measured by APFD (Average Percentage of Faults Detected), we show that coverage-guided test generation exposes distinct retrieval faults substantially earlier than random selection. CC improves APFD by 10\% to 25\% across datasets, indicating earlier discovery of diverse failure modes, showing that CC captures aspects of retrieval diversity that are directly relevant to effective testing.

In summary, this paper makes the following contributions:
\begin{itemize}
    \item We introduce \textit{Chunk Coverage (CC)}, an oracle-independent adequacy criterion for testing the retrieval component of RAG systems at the test-suite level.
    \item We present a coverage-guided test selection and generation process that systematically exercises under-tested regions of the retrieval space.
    \item We empirically demonstrate, across realistic RAG scenarios, that higher CC leads to faster coverage growth and earlier detection of distinct retrieval faults, outperforming random and redundancy-biased baselines.
\end{itemize}

\new{The remainder covers background and CC (Sections~\ref{sec:background}--\ref{sec:chunk_coverage}), evaluation (Sections~\ref{sec:experimental_setup}--\ref{sec:results}), related work and threats to validity (Sections~\ref{sec:related_work}--\ref{sec:threats}), and conclusions (Section~\ref{sec:conclusion}).}

% !TEX root = ../main.tex

\section{Background}
\label{sec:background}

This section introduces the fundamentals of Retrieval-Augmented Generation (RAG) systems and reviews existing evaluation metrics that assess their retrieval and generation behaviour.

\subsection{Retrieval-Augmented Generation (RAG)}
\label{sec:background_rag}

\begin{figure}[t]
  \centering
  \begin{subfigure}{\textwidth}
     \centering
    \includegraphics[width=0.8\linewidth]{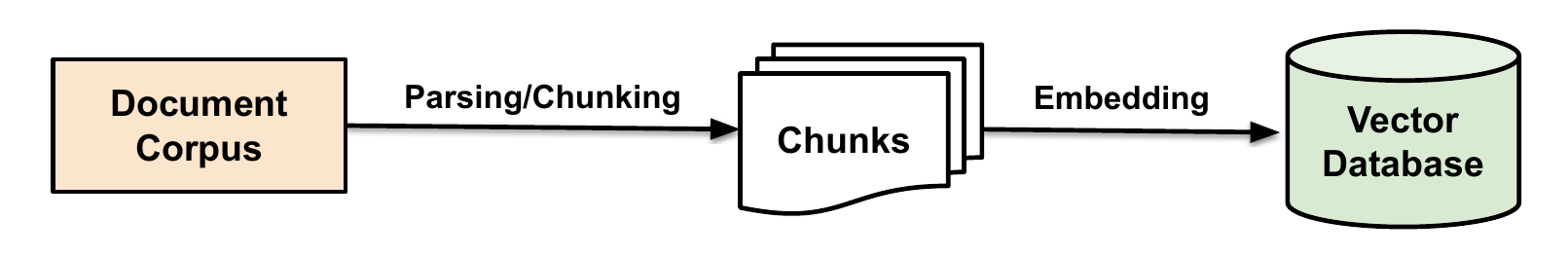}
    \caption{Document Indexing (Offline)}\label{fig:rag_indexing}
  \end{subfigure}
  \begin{subfigure}{\textwidth}
     \centering
    \includegraphics[width=0.8\linewidth]{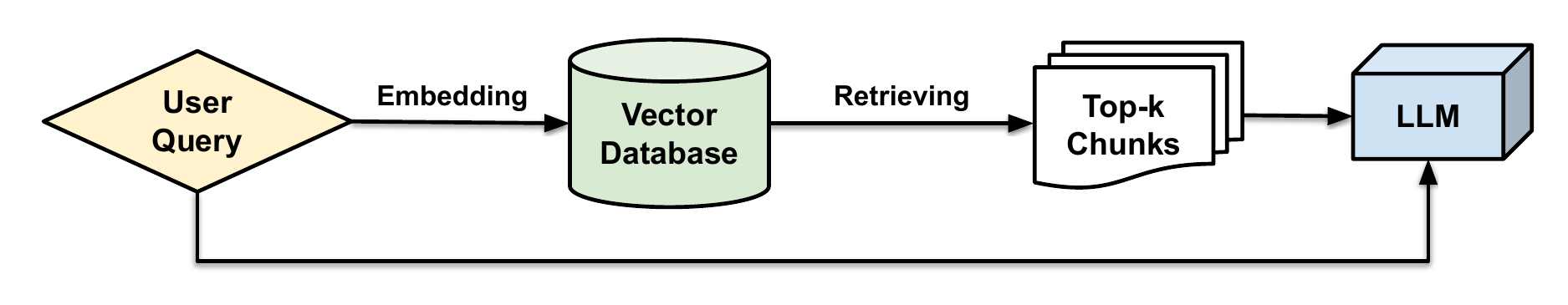}
    \caption{Output Generation (Online)}\label{fig:rag_generating}
  \end{subfigure}
  \caption{Overview of RAG Procedure}\label{fig:rag}
\end{figure}

RAG is a widely adopted architectural pattern for LLMs that augments text generation with external documents retrieved at inference time. Rather than relying solely on parametric knowledge encoded during training of LLM, a RAG system incorporates information from an explicit document corpus to ground model outputs. Figure~\ref{fig:rag} illustrates the standard RAG procedure, which can be decomposed into two phases: document indexing and output generation. %\samuele{If we have space at the end, we can expand a little, specifying that the research on RAG focused a lot on making these systems flexible to parse from a lot of different types of documents and to make the pipeline extremely modular, and this is one of the main reason of success of frameworks such as Langchain and LlamaIndex} \han{Let me see but we gotta cut some parts now.}

 \paragraph{Document indexing.}
As shown in Figure~\ref{fig:rag_indexing}, the document indexing phase is performed offline and prepares the corpus for retrieval. The documents are first parsed and segmented into smaller textual units, commonly referred to as \textit{chunks}. Each chunk is then embedded into a high-dimensional \textit{vector space} using a text embedding model. The resulting vectors are stored in a vector database or index that supports similarity-based retrieval. This embedding space defines the retrieval behaviour of the system: chunks that are close in the vector space are more likely to be retrieved together.

\paragraph{Output generation.}
Figure~\ref{fig:rag_generating} depicts the online inference phase. Given an input query, the query text is embedded using the same embedding model and used to retrieve the top-$k$ most similar chunks from the vector database according to a distance or similarity measure (e.g., cosine similarity). This retrieval step determines which parts of the corpus are selected for use during generation and constitutes the primary decision-making component of a RAG system. The retrieved chunks are then incorporated into the final prompt provided to the LLM, typically via concatenation or structured formatting. The LLM produces the system’s response conditioned on both the original query and the retrieved context.

\subsection{Evaluation Metrics for RAG Systems}
\label{sec:background_rag_eval}

The evaluation of RAG systems has predominantly focused on assessing the quality of retrieved context and generated responses at the level of individual queries. Existing RAG evaluation metrics are primarily designed for performance assessment and benchmarking, rather than for analysing the adequacy of a test set (i.e., a collection of queries) or guiding test generation.

A prominent line of work proposes automated, reference-free evaluation metrics that approximate human judgements using LLMs. RAGAS~\cite{es2024ragas} introduces a suite of such metrics that decompose RAG evaluation into multiple dimensions, spanning both retrieval and generation. At the retrieval level, metrics such as \textit{Context Precision} and \textit{Context Relevance} assess how well the retrieved chunks align with the information needs expressed by a query. At the generation level, metrics including \textit{Response Relevancy}, \textit{Faithfulness}, and \textit{Response Groundedness} evaluate whether the generated answer addresses the query and whether its claims are supported by the retrieved context. These metrics are computed independently for each query--response pair and are typically aggregated across a dataset to obtain overall performance scores.

Formally, for a test query $q \in T$, let $R_k(q) \subseteq \mathcal{C}$ denote the set of top-$k$ retrieved chunks, and let $a(q)$ be the generated response. At the retrieval level, \textit{Context Precision} measures the fraction of retrieved chunks judged to be relevant to answering the query:
\begin{equation*}
\text{Context Precision}(q) =
    \frac{|\{\, c \in R_k(q) \mid \mathrm{rel}(c,q) = 1 \,\}|}{k}
\end{equation*}
where $\mathrm{rel}(c,q)\in\{0,1\}$ indicates whether chunk $c$ is relevant to query $q$, as assessed by an LLM-as-a-judge.

Closely related, \textit{Context Relevance} is computed at the sentence level. For a query $q$, the retrieved context is first concatenated, and an auxiliary LLM (distinct from the LLM under test) is used to identify the subset of sentences $S_{\mathrm{ext}}(q)$ that are necessary to answer the query. The metric is then defined as:
\begin{equation*}
\text{Context Relevance}(q) = \frac{|S_{\mathrm{ext}}(q)|}{|S(c(q))|}
\end{equation*}
where $S(c(q))$ denotes the set of all sentences in the retrieved context.

At the generation level, \textit{Response Relevancy} estimates how well the generated answer addresses the original query. This metric is computed by prompting an auxiliary LLM to generate a set of questions $\{q_i\}_{i=1}^n$ that are answerable by the generated answer $a(q)$, and then averaging their semantic similarity to the original query:
\begin{equation*}
\text{Response Relevancy}(q) = \frac{1}{n}\sum_{i=1}^{n} \mathrm{sim}(q,q_i)
\end{equation*}

Finally, \textit{Faithfulness} evaluates whether the content of the generated answer is supported by the retrieved context. The answer is first decomposed into a set of atomic statements $S(a(q))$, and the metric is defined as the proportion of statements that are supported by the retrieved chunks:
\begin{equation*}
\text{Faithfulness}(q) =
\frac{|\{\, s \in S(a(q)) \mid s \text{ is supported by } R_k(q) \,\}|}{|S(a(q))|}
\end{equation*}

\new{\paragraph{Illustrative example.} Consider a financial RAG system using the prompt template ``Context: [retrieved chunks]; Question: [query]; Answer using only the context.'' For the query ``By what percentage did revenue increase from 2023 to 2024?'', its top-3 retriever returns $c_1$: ``Revenue was \$8 million in 2023'', $c_2$: ``Revenue was \$10 million in 2024'', and the irrelevant $c_3$: ``The company is headquartered in Zurich.'' The system answers ``Revenue increased by 25\%.'' If an LLM judge marks $c_1$ and $c_2$ as relevant, Context Precision is $2/3=0.67$; because each chunk contains one sentence and two are necessary, Context Relevance is also $2/3=0.67$. If the auxiliary LLM generates two questions with similarities 0.94 and 0.90 to the original query, Response Relevancy is $(0.94+0.90)/2=0.92$, while the answer's single claim is supported by $c_1$ and $c_2$, yielding Faithfulness of 1.0. Yet, if a 100-chunk corpus is tested only by queries that retrieve these same three chunks, the suite exercises just $3/100=3\%$ of the corpus, illustrating why high per-query scores do not establish suite-level test adequacy.}

Similar evaluation goals are pursued by ARES~\cite{saadfalcon2023ares}, which proposes an automated framework for assessing RAG systems by judging the correctness and grounding of generated answers with respect to retrieved documents. ARES focuses on identifying unsupported or hallucinated claims by comparing generated responses against retrieved evidence, again relying on LLM-based judgements in the absence of explicit ground truth annotations.

While these metrics are well-suited for performance evaluation and system comparison, they provide limited insight into how a collection of tests collectively exercises the retrieval behaviour of a RAG system. Moreover, due to the common absence of ground-truth relevance and grounding annotations, existing metrics typically rely on an LLM-as-a-judge to approximate human assessments, which introduces additional subjectivity and uncertainty into the evaluation. Conceptually, these metrics are defined at the level of individual query--response pairs and are primarily intended to be averaged across a dataset, rather than to characterise properties of the query \textit{set} itself. For instance, consider a RAG system deployed over hundreds of pages of internal technical documentation or regulatory reports. In such settings, developers must validate not only that individual queries retrieve relevant context, but also that the test suite as a whole exercises the retriever across the corpus, rather than repeatedly retrieving the same small subset of frequently accessed documents while leaving large portions of the corpus \textit{untested}. This motivates the need for test adequacy metrics that characterise how thoroughly a RAG system’s retrieval behaviour is exercised by a \textit{test set}, independently of per-query retrieval quality or answer correctness.

% !TEX root = ../main.tex

% \han{Figure for RAG system and CC.} \\
% \han{Need to specify what we're going to test (testing target): two parts: embedding model and similarity algorithm} \\
% \han{Highlight that CC does not reveal the faults (i.e., retrieval error) like other RAG evaluation metrics. By the way, to do so, it needs a ground truth retrieved chunk :p}
% \han{Need better scenario where the developers want to have higher coverage.}
% \han{This is analogous to: code coverage testing a program’s execution, not benchmarking which compiler is better.}

% Adequacy of a test suite under a fixed retrieval configuration
% “Does this test suite exercise the retrieval behaviour induced by embedding X + DB Y?”
% “Does the same test suite exercise different parts of the retrieval space when the vector DB changes?”

\section{Chunk Coverage for Testing RAG Systems}
\label{sec:chunk_coverage}

% In many practical deployments of RAG, the external document corpus constitutes the primary source of task-relevant information. The LLM is expected to generate correct outputs by conditioning on retrieved documents, rather than relying on its own parametric or open-domain knowledge. Typical examples include question answering over proprietary technical documentation, software repositories, scientific articles, legal texts, or internal knowledge bases. In such settings, the behaviour of the retrieval component plays a central role in determining which information is accessible to the system at inference time.

% From a testing perspective, this raises the question of whether a given test suite sufficiently exercises the retrieval behaviour of a RAG system. Existing evaluation practices predominantly assess retrieval quality or answer correctness on a per-query basis. Metrics such as context precision and recall quantify whether the retrieved chunks for a given query are relevant, typically by comparing them against ground-truth relevance annotations. While effective for performance evaluation, these metrics provide limited insight into how a collection of tests collectively exercises the retriever. In particular, they do not characterise which parts of the document corpus or embedding space have been explored by the test suite, and which parts remain untested.

This work focuses on testing the retrieval component of RAG systems. Concretely, we consider retrieval behaviour as determined by two core elements: (1) the embedding model that maps queries and chunks into a shared vector space, and (2) the vector database together with its retrieval configuration (e.g., similarity metric, indexing strategy, and top-$k$ policy). Together, these elements define the regions of the document corpus that are reachable by the retriever under a given configuration.

To characterise how thoroughly a test suite exercises this retrieval behaviour, we introduce \textit{Chunk Coverage} (CC), a structural adequacy criterion for RAG systems. Given a test suite $T$, CC measures the proportion of chunks in the corpus that are retrieved at least once across all test executions:  
\begin{equation*}
\label{eq:cc}
\mathrm{CC}(T) = 
\frac{\left| \bigcup_{q \in T} R_k(q) \right|}
     {|\mathcal{C}|}
\end{equation*}
where $T$ is a test suite (set of queries), $\mathcal{C}$ is the set of all chunks in the corpus, and $R_k(q) \subseteq \mathcal{C}$ denotes the set of chunks retrieved for query $q$, considering the top-$k$ results.

% $\mathrm{CC}(T) = \dfrac{\left| \bigcup_{q \in T} R_k(q) \right|}{|\mathcal{C}|}$
By construction, CC characterises retrieval behaviour independently of downstream generation quality and does not require ground-truth relevance judgements. Importantly, CC is not designed to detect retrieval faults or to assess whether retrieved chunks are correct or relevant for a given query. Identifying such errors inherently requires a test oracle, such as ground-truth relevance annotations, and is the objective of existing RAG evaluation metrics. CC instead serves a complementary purpose: it assesses test adequacy by indicating which regions of the retrieval space induced by the embedding model and retrieval configuration have been exercised by a test suite. Unlike existing RAG evaluation metrics, CC captures the \textit{global} exploration of the retrieval space induced by multiple queries. As an oracle-independent criterion, CC can be optimised, targeted, or maximised during test generation to systematically exercise under-tested regions of the retriever’s behaviour space.

\begin{figure}[t]
    \centering
    \includegraphics[width=\textwidth]{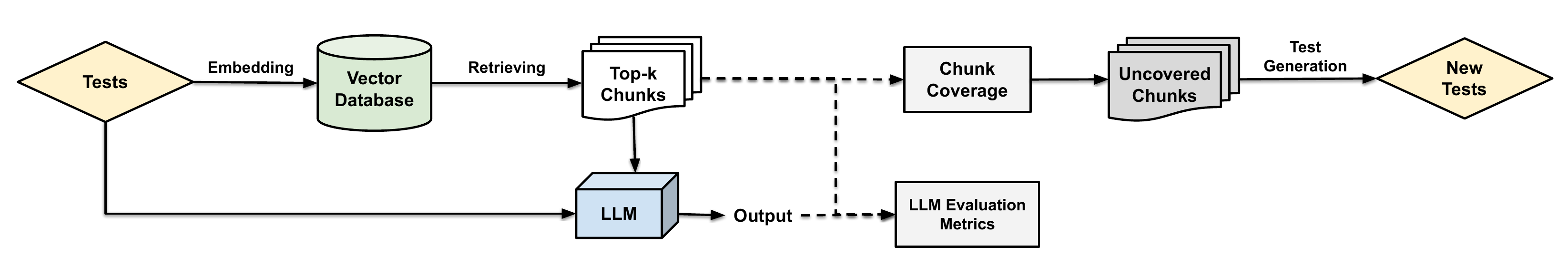}
    \caption{Test Generation with Chunk Coverage}
    \label{fig:testgen}
\end{figure}

\subsection{Interpretation and Applicability of Chunk Coverage}

From a software testing perspective, CC plays a role analogous to code coverage in traditional programs. It abstracts away from output correctness and does not aim to compare alternative retrieval implementations or configurations. Instead, CC characterises whether the behavioural space of a fixed retrieval configuration has been sufficiently exercised by a test suite, thereby isolating the retriever as an explicit test target and enabling focused, coverage-guided testing of RAG systems.

As with traditional code coverage, not all uncovered elements necessarily warrant additional testing. In conventional programs, certain code regions, such as logging statements, may be intentionally deprioritised or excluded from coverage goals, even though they contribute to the overall coverage metric. Similarly, in RAG systems, some document chunks may be out of scope for the intended task, contain auxiliary or boilerplate information, or be deliberately ignored by developers. CC makes such unexercised regions explicit, allowing developers to decide which portions of the retrieval space are worth further testing, rather than implying that full (100\%) coverage is always desirable.

It is worth noting that CC itself makes no assumption about whether the LLM relies on the retrieved documents to produce correct answers. CC purely characterises the behaviour of the retrieval component by measuring which parts of the document corpus are exercised by a test suite. Nevertheless, CC is most informative in RAG settings where the external document corpus constitutes a necessary source of task-relevant information. In open-knowledge tasks, where an LLM can often answer queries correctly without relying on retrieved documents, low coverage does not necessarily indicate inadequate testing, and full coverage is neither required nor expected. In such settings, CC should therefore be interpreted as a characterisation of retriever behaviour rather than as a measure of test adequacy. Accordingly, in our experiments, we focus on realistic deployment scenarios in which retrieval is essential for correct system behaviour, and CC can serve as a meaningful adequacy criterion.

\subsection{Test Generation with Chunk Coverage}
\label{sec:testgen_with_cc}

\begin{algorithm}[H]
\caption{Coverage-guided test generation with Chunk Coverage}
\label{alg:cc-testgen}

\begin{algorithmic}[1]
\small
\Require Chunk corpus $\mathcal{C}$, retriever $\mathcal{R}$, $k$, initial test suite $T_0$, budget $B$, target coverage $\tau$
\Ensure Test suite $T$
\State $T \gets T_0$ \label{line:initT}
\State $E \gets \emptyset$ \Comment{Retrieved (exercised) chunks} \label{line:initE}
\While{\textsc{BudgetNotExpired}($B$)}
    \ForAll{$q \in T$ \new{\textbf{ while } \textsc{BudgetNotExpired}($B$)}} \label{line:forq}
        \State $R \gets \mathcal{R}_k(q)$ \Comment{Execute retrieval}
        \State \new{$B \gets B - 1$} \Comment{\new{Consume one query execution}} \label{line:budget}
        \State $E \gets E \cup R$ \label{line:updateE}
        % \State \textsc{OptionallyEvaluate}$(q, R)$ \Comment{LLM-based RAG metrics} \label{line:eval}
    \EndFor
    \State $\mathrm{CC}(T) \gets \frac{|E|}{|\mathcal{C}|}$ \label{line:cc}
    \If{$\mathrm{CC}(T) \ge \tau$} \label{line:stop1}
        \State \textbf{break} \label{line:stop2}
    \EndIf
    \State $U \gets \mathcal{C} \setminus E$ \Comment{Uncovered chunks} \label{line:uncovered}
    \ForAll{$c \in U$} \label{line:forc}
        \State $\mathcal{Q}_c \gets \textsc{GenerateQueries}(c)$ \Comment{e.g., via an auxiliary LLM} \label{line:genQ}
        \State $T \gets T \cup \mathcal{Q}_c$ \label{line:add}
    \EndFor
\EndWhile
\State \Return $T$
\end{algorithmic}
\end{algorithm}

CC enables a coverage-guided approach to test generation for RAG systems by providing explicit feedback on which parts of the retrieval space have been exercised by a test suite. Algorithm~\ref{alg:cc-testgen} formalises it: the algorithm takes as input a document corpus $\mathcal{C}$, a RAG retriever $\mathcal{R}$ (including the embedding model and vector-store configuration), a retrieval parameter $k$, an initial test suite $T_0$ (which may be empty), a testing budget $B$, and a target coverage threshold $\tau$. Figure~\ref{fig:testgen} illustrates the overall process.

Test generation starts from the initial test suite and an empty set of exercised chunks (Algorithm~\ref{alg:cc-testgen}, Lines~\ref{line:initT}--\ref{line:initE}). Each test query is executed against the RAG system, triggering embedding, retrieval from the vector database, and the selection of top-$k$ chunks (Lines~\ref{line:forq}--\ref{line:updateE}). \new{Each retrieval consumes one unit of $B$, and the inner loop stops when no budget remains (Lines~\ref{line:forq}--\ref{line:budget}).} The retrieved chunks are recorded and aggregated to compute and update the current CC value (Line~\ref{line:cc}). Optionally, during this execution phase, existing RAG evaluation metrics that require additional LLM inference may be computed. Such evaluations are orthogonal to CC and are not required for coverage computation or test generation. In particular, Algorithm~\ref{alg:cc-testgen} does not use these metrics as feedback signal to guide test generation.

Based on the updated coverage information, the algorithm identifies chunks that have not yet been retrieved (Line~\ref{line:uncovered}). These uncovered chunks correspond to under-exercised regions of the retrieval space induced by the embedding model and vector database. For each uncovered chunk, the algorithm generates one or more new test queries with the explicit goal of retrieving that chunk (Lines~\ref{line:forc}--\ref{line:genQ}). One possible instantiation of this step is to employ an auxiliary LLM that takes the content of an uncovered chunk as input and synthesises queries intended to retrieve it. The newly generated queries are added to the test suite (Line~\ref{line:add}), and the process iterates.

This feedback-directed loop continues until the testing budget is exhausted or the target coverage threshold is reached (Lines~\ref{line:stop1}--\ref{line:stop2}). Importantly, CC-guided test generation relies solely on observable retrieval behaviour and does not require ground-truth relevance annotations or expected answers. As a result, it remains applicable in settings where defining precise test oracles is difficult or infeasible.

From a software testing perspective, this process mirrors coverage-guided testing in traditional software systems, where execution feedback is used to steer test generation towards untested program regions. In the context of RAG systems, CC provides an analogous feedback signal at the level of retrieval behaviour, enabling systematic and targeted exploration of the retriever independently of downstream generation quality.

% !TEX root = ../main.tex

\section{Experimental Setup}
\label{sec:experimental_setup}

\subsection{Research Questions}

We structure our empirical study around the following three research questions.

\textbf{RQ1.} \textit{How does Chunk Coverage relate to retrieval diversity at the test-suite level?}
We examine whether CC aligns with established notions of diversity. We construct 100 test suites of equal size ($n \in \{30, 50, 100\}$) using uniform random query selection (i.e., a strategy that does not rely on CC for guidance) and analyse the relationship between CC and suite-level diversity metrics: the normalized entropy, Gini coefficient of chunk retrieval frequencies, and the average pairwise Jaccard overlap of retrieved chunk sets. This analysis validates whether CC behaves consistently with intuitive expectations of behavioural exploration in the retrieval space, i.e., greater CC is expected to be associated with higher diversity of the test queries.

\textbf{RQ2.} \textit{How effective is Chunk Coverage at guiding the generation/selection of tests that systematically explore the retrieval behaviour?} 
We investigate whether CC can serve as an effective feedback signal for constructing test suites that progressively expand coverage of the retrieval space. We compare CC-guided test generation/selection (see Section~\ref{sec:scenarios_and_test_generations} for details) against both baselines in terms of coverage growth and achieved CC under a fixed testing budget, assessing whether CC provides principled guidance for achieving faster and more comprehensive exploration of retrieval behaviour.

\textbf{RQ3.} \textit{How does Chunk Coverage relate to testing effectiveness?} 
We investigate whether higher CC leads to more effective testing by exposing distinct faults earlier under a limited testing budget, even though CC itself is not designed to detect faults. As test suites are incrementally augmented, we analyse how increasing CC affects the occurrence and ordering of failures, as identified through established RAG evaluation metrics capturing retrieval quality and answer grounding. Following the definitions in Section~\ref{sec:failures_and_faults}, extreme values of these metrics are interpreted as indicators of observable failure symptoms, while faults correspond to distinct underlying retrieval behaviours that manifest through such failures. Testing effectiveness is quantified using APFD (Average Percentage of Faults Detected), which measures how early distinct faults are exposed as the test suite grows. This allows us to assess whether test suites achieving higher CC also detect faults earlier, with the key benefit that prioritisation by CC does not rely on explicit test oracles.

% RQ3. Cost–benefit and diminishing returns of CC??
% Cost experiment?
% Redundancy ratio (how many tests add nothing new)
% Coverage distribution across chunks (optional but nice) e.g., entropy or Gini over retrieval frequencies.

% Failures found vs budget curve
% Can note that RAGAS metrics are not useful for test-suite level.
% Spearman correlation between CC and each RAGAS metric; Low correlation → CC captures something different (good for novelty).
% Can think of multi-objective variants that combines CC and RAGAS metrics.

\begin{table}[t]
\small
\centering
\caption{Descriptive characteristics of the benchmarks used in our evaluation.}
\label{tab:dataset_characteristics}
\begin{tabular}{lll}
\toprule
\textbf{Characteristic} & \textbf{MIMIC-IV} & \textbf{T2-RAGBench} \\
\midrule
Application domain & Clinical decision making & Financial question answering \\
Source documents & Patient cases & Financial reports and filings \\
Modalities & Textual clinical records & Text and tables \\
Granularity of documents & Per-patient case & Per document / page \\
Predefined test queries & None & Yes \\
Ground-truth answers & None & Yes \\
Ground-truth relevance & None & Page-level \\
Chunk-level relevance labels & None & None \\
Intended reliance on retrieval & Essential & Essential \\
\bottomrule
\end{tabular}
\end{table}

\begin{table}[t]
\centering
\caption{The size of the induced chunk space, the number of available test queries, and the maximum attainable Chunk Coverage for each dataset.}
\label{tab:datasets-specs}
\small
\begin{tabular}{llrrr}
\toprule
\textbf{Domain} & \textbf{Dataset} & \textbf{\# Chunks} & \textbf{\# Queries} & \textbf{Max CC (\%)} \\
\midrule
Medical & MIMIC-IV~\cite{hager_mimic-iv-ext_nodate} & 2{,}160 & 1{,}709 & 25.42 \\
Financial & ConvFinQA~\cite{chen2021finqa} & 13{,}444 & 3{,}454 & 13.28 \\
Financial & FinQA~\cite{chen2021finqa} & 16{,}940 & 6{,}251 & 16.89 \\
Financial & TAT-DQA~\cite{zhu2022towards} & 16{,}552 & 9{,}059 & 24.02 \\
Financial & VQAonBD~\cite{raja2023icdar}  & 5{,}722 & 9{,}813 & 38.33 \\
\bottomrule
\end{tabular}
\end{table}

\subsection{Datasets and RAG Settings}

Tables~\ref{tab:dataset_characteristics} and~\ref{tab:datasets-specs} summarise the benchmarks used in our evaluation from two complementary perspectives. Table~\ref{tab:dataset_characteristics} provides a high-level characterisation of the application domains, data modalities, and available supervision. Table~\ref{tab:datasets-specs} focuses on properties that are directly relevant to our study of retrieval adequacy, including the size of the induced chunk space, the number of test queries, and the maximum attainable CC in our study setting. In total, we consider five datasets from two high-stakes application domains in which correct system behaviour depends on retrieving external documents rather than on the parametric knowledge of the model alone, as detailed below. 

% \new{More specifically, we required document-grounded question answering and a document structure suitable for chunk-level evaluation. These criteria led us to include all four T2-RAGBench datasets, which span textual and tabular financial evidence, and MIMIC-IV, the only publicly available clinical RAG benchmark we identified with suitable patient-level documents.}

\subsubsection{Clinical Decision-making over Patient Records}

The clinical decision-making scenario is based on the MIMIC-IV Clinical Decision Making dataset~\cite{hager_mimic-iv-ext_nodate}, a curated and task-oriented derivative of the MIMIC-IV electronic health record database~\cite{johnson2023mimic}. The dataset consists of individual patient cases derived from real hospital admissions and covers four abdominal pathologies: appendicitis, cholecystitis, diverticulitis, and pancreatitis. Each patient case aggregates multiple clinical modalities, including history of present illness, physical examination findings, laboratory and microbiology results, radiology reports, and recorded procedures.

\new{To prevent trivial solutions based on explicit labels, it removed or anonymised diagnostic mentions and excluded patient cases missing primary modalities.} As a result, answering clinical queries requires synthesising evidence across multiple pieces of retrieved patient-specific information, rather than recalling diagnoses or relying on general medical knowledge encoded in the LLM. Each patient case is treated as a separate document and segmented into chunks that represent distinct clinical evidence units, which are embedded and indexed in a vector database for retrieval. Note that the dataset does not provide predefined test queries, ground-truth answers, or relevance annotations (see Table~\ref{tab:dataset_characteristics}). Test queries are therefore generated as part of our experimental procedure with the guidance of CC and correspond to clinical decision-making questions posed to an RAG-based LLM agent.

\subsubsection{\new{Financial QA over Heterogeneous Documents}}

The financial question-answering scenario is based on T2-RAGBench~\cite{strich2025t}, a benchmark designed to evaluate RAG over large collections of financial documents containing both textual and tabular information. T2-RAGBench aggregates four datasets: ConvFinQA~\cite{chen2021finqa}, FinQA~\cite{chen2021finqa}, TAT-DQA~\cite{zhu2022towards}, and VQAonBD~\cite{raja2023icdar} that cover heterogeneous financial sources such as annual reports, regulatory filings, and financial statements.

Unlike MIMIC-IV, all four datasets in T2-RAGBench provide predefined test queries and verified answers, with relevance defined at the page-level (see Table~\ref{tab:dataset_characteristics}). Test queries in these datasets typically require multi-step reasoning over retrieved evidence, such as locating specific tables or passages, extracting numerical values, and performing comparisons or arithmetic operations. Each query is associated with a verified answer and supporting context at the document-page level. Correct system behaviour therefore depends on retrieving the appropriate page containing the relevant evidence.

In our RAG setup, documents or individual pages are segmented into smaller chunks and embedded in a vector database for retrieval. Although relevance annotations are available at the page level, no supervision exists at the finer-grained chunk level. As a result, even when all available queries are executed, only a subset of the induced chunk space can be exercised. This is reflected in the maximum attainable CC reported in Table~\ref{tab:datasets-specs}, which remains strictly below 100\%.

\subsection{Failures and Faults in RAG Systems}
\label{sec:failures_and_faults}

To study testing effectiveness, we require operational definitions of failures and faults. In line with common testing practice, we treat failures as observable deviations from desired behaviour, and faults as their underlying root causes. For RAG systems, we define failures in terms of violations of intended RAG behaviour, as reflected by established evaluation dimensions. A failure is defined at the level of an individual test query: given a query \(q\), we consider the RAG system to have failed if its observed behaviour exhibits a clear violation along at least one evaluation dimension, such as relevance, grounding, or retrieval alignment.

Operationally, we detect such violations using existing RAG evaluation metrics. Let \(M\) denote the set of metrics used in our study, e.g., response relevancy, context precision, and faithfulness. Extremely low metric values indicate pronounced deviations from the intended behaviour captured by the respective metric. In our experiments, we treat the lowest attainable metric value (i.e., zero) as a concrete indicator of failure, as it corresponds to cases where the evaluated property is entirely absent. This allows us to capture a range of failure symptoms. For example, a context relevance value at the extreme lower bound indicates that the retrieved context is misaligned with the user input, pointing to a retrieval failure. Conversely, extreme values for faithfulness indicate that the generated answer is unsupported by the retrieved context, suggesting a failure in grounding or generation. In all cases, the metrics serve as indicators of observable failure symptoms rather than as definitions of failure.

While failures capture observable deviations in behaviour, they do not directly correspond to \textit{distinct} underlying problems. As in conventional software testing, e.g., fuzz testing, multiple test inputs may trigger the same underlying fault. Counting each such occurrence separately would therefore overestimate testing effectiveness. Consequently, we define a fault as an underlying root cause in the RAG system that gives rise to one or more failures.
In this work, we focus on faults rooted in the retrieval component, as retrieval behaviour fundamentally constrains the information available to the generator and is directly actionable for system developers. Typical retrieval faults may arise from the embedding model used to map queries and chunks into a shared vector space, from the vector database and its indexing or similarity configuration, or from retrieval parameters such as the top-$k$ policy. Identifying faults at this level provides concrete guidance on where corrective actions, such as re-embedding, re-indexing, or configuration changes, should be applied.

Two failing queries are considered to correspond to the same fault if they induce the same or highly overlapping retrieval behaviour. Formally, let \( R_k(q) \) denote the set of chunks retrieved for a query \( q \). We measure similarity between retrieval behaviours using the Jaccard similarity of the retrieved chunk sets, treating the retrieved chunks as an unordered set. If the Jaccard similarity between \( R_k(q_1) \) and \( R_k(q_2) \) exceeds a threshold \( \tau \) (0.8 in our study), the corresponding failures are attributed to the same fault and counted once. By grouping failures in this way, each identified fault corresponds to a distinct region of the retrieval space that is systematically mishandled under the current configuration. In RQ3 (Section~\ref{sec:rq3}), we use this definition to study how increasing Chunk Coverage relates to the discovery of new faults and how such faults are reflected in existing RAG evaluation metrics. \new{We chose $\tau=0.8$ as a conservative criterion for grouping only failures with strongly overlapping retrieval behaviour. To assess sensitivity to this choice, we repeated RQ3 with $\tau \in \{0.6, 0.7, 0.8, 0.9, 1.0\}$: the APFD of the coverage-guided strategy varied by at most 0.1 percentage points, and the ranking CC $>$ Random $>$ Overlap-biased held in 140 of 145 (96.6\%) dataset--metric--threshold combinations. The five exceptions were the MIMIC-IV Response Groundedness cases, for which Random slightly outperformed CC at every threshold; thus, our conclusions are insensitive to the threshold choice.}

\subsection{Scenarios and Test Generations/Selections}
\label{sec:scenarios_and_test_generations}

Our study distinguishes between how test queries are obtained and how test suites are constructed and evaluated under different scenarios. This separation allows us to analyse the effect of coverage guidance on retrieval exploration while controlling for the source of test inputs.
At present, there is no consolidated, automated test generation framework for RAG systems. As a result, test queries must either be drawn from existing, manually defined query pools or generated using auxiliary LLMs, depending on dataset availability.

\subsubsection{Evaluation Scenarios}

We consider three evaluation scenarios that differ in how test suites are incrementally constructed from a given pool of queries. First, in the \textit{coverage-guided} (i.e., CC-guided) scenario, test queries are generated/selected to maximise incremental CC. At each step, queries that retrieve previously unexercised chunks are prioritised, with the explicit goal of systematically expanding coverage of the induced retrieval space. Second, the \textit{overlap-biased} scenario represents a worst case from the perspective of CC. Test queries are selected to maximise overlap with already exercised retrieval behaviour, prioritising queries whose retrieved chunk sets strongly intersect with those observed so far. This deliberately limits exploration of new retrieval regions and serves as a conservative baseline. Third, in the \textit{random} scenario, test queries are selected uniformly at random from the available pool, without regard to coverage or overlap. This reflects common practice in the absence of explicit adequacy criteria.

\new{We note that we do not use entropy or the Gini coefficient as guidance signals because they are aggregate suite-level statistics rather than actionable per-query criteria: unlike CC, neither identifies which chunks remain untested. Turning them into selection strategies would require designing new greedy techniques and recomputing the suite-level retrieval distribution for every candidate at each step; moreover, both metrics can change non-monotonically as tests are added and thus provide neither a coverage target nor a stopping condition.}

All three scenarios operate on the same underlying query pool and differ only in the query selection strategy, allowing us to isolate the effect of coverage guidance. The way in which the initial pool of test queries is obtained differs across datasets and is discussed next.

\subsubsection{Simulated Evaluation Using Existing Queries (Selection)}

For T2-RAGBench, a large pool of predefined test queries is available. In this case, we do not generate additional queries. Instead, we simulate the three evaluation scenarios by incrementally selecting queries from the existing pool according to the coverage-guided, overlap-biased, and random strategies. This enables controlled comparisons without introducing additional variability from LLM-generated inputs.

\subsubsection{LLM-assisted Test Generation}

In contrast, MIMIC-IV does not provide predefined test queries, nor does the literature offer established guidance for generating queries in this RAG setting. We therefore rely on LLM-assisted test generation to obtain a pool of test queries, as illustrated in Section~\ref{sec:testgen_with_cc}. Specifically, we apply Algorithm~\ref{alg:cc-testgen}, which uses an auxiliary LLM (Line~\ref{line:genQ}) to generate candidate queries targeting uncovered chunks identified through CC. The auxiliary LLM is instructed to generate clinically plausible questions that require the information in those chunks to be answered.

Once this pool is constructed, we treat it as fixed and simulate the same three evaluation scenarios by selecting queries according to the respective strategies. In this way, CC is used to guide the generation of candidate queries, while all evaluation scenarios are compared under identical conditions. \new{Thus, all three strategies operate on the same fixed, CC-generated query pool and differ only in their query ordering.}

\subsection{Configurations}

In the study, we retrieve the top-$k$ chunks with $k=3$. For the LLM-assisted test generation scenario, the budget is set to the size of the chunk space, i.e., with the MIMIC-IV, this allows us to target each uncovered chunk with a test generation attempt. Experiments involving non-deterministic behaviour, such as random query selection, are repeated ten times. We report average results, and for figures, we include confidence intervals to reflect variability across runs.

\new{For both answer generation and LLM-as-a-judge evaluation, we use OpenAI's pinned GPT-4o-mini\footnote{gpt-4o-mini-2024-07-18} with temperature 1, together with \texttt{text-embedding-3-small}\footnote{\url{https://platform.openai.com/docs/models/text-embedding-3-small}} for embeddings.}
The prompts used in the RAG systems are provided in our artefact (see Section~\ref{sec:data}). All experiments are conducted on a machine equipped with an Apple M3 Max and 36GB of RAM, running Tahoe 26.2.

% !TEX root = ../main.tex

\section{Results}
\label{sec:results}

This section presents the empirical results for the three research questions.

\begin{figure}[t]
    \centering

    \begin{subfigure}[b]{0.32\textwidth}
        \centering
        \includegraphics[width=\linewidth]{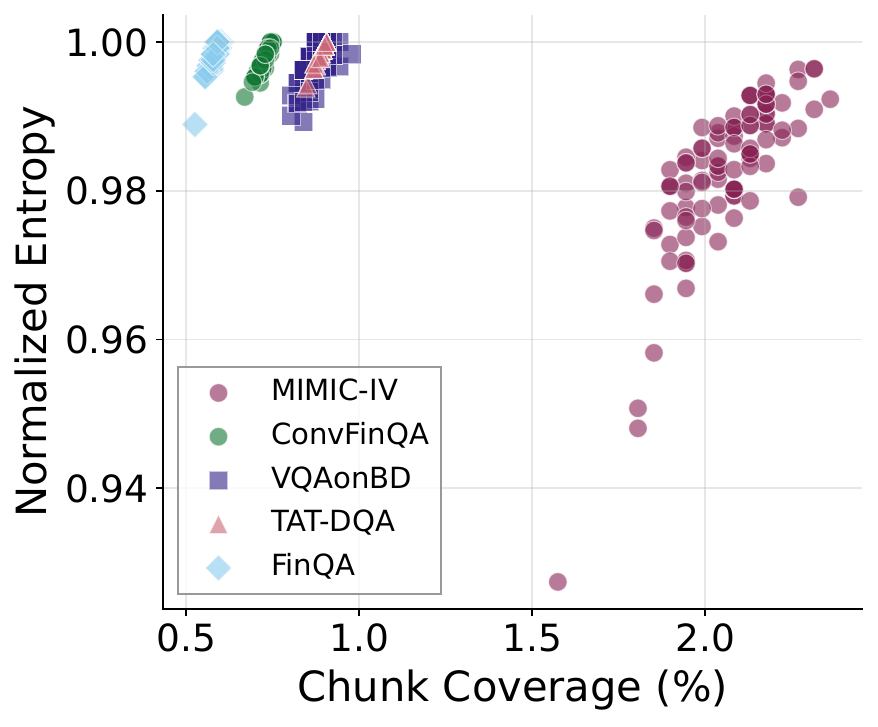}
        \caption{Normalised Entropy} \label{fig:rq1-scatter-entropy}
    \end{subfigure}
    \hfill 
    \begin{subfigure}[b]{0.32\textwidth}
        \centering
        \includegraphics[width=\linewidth]{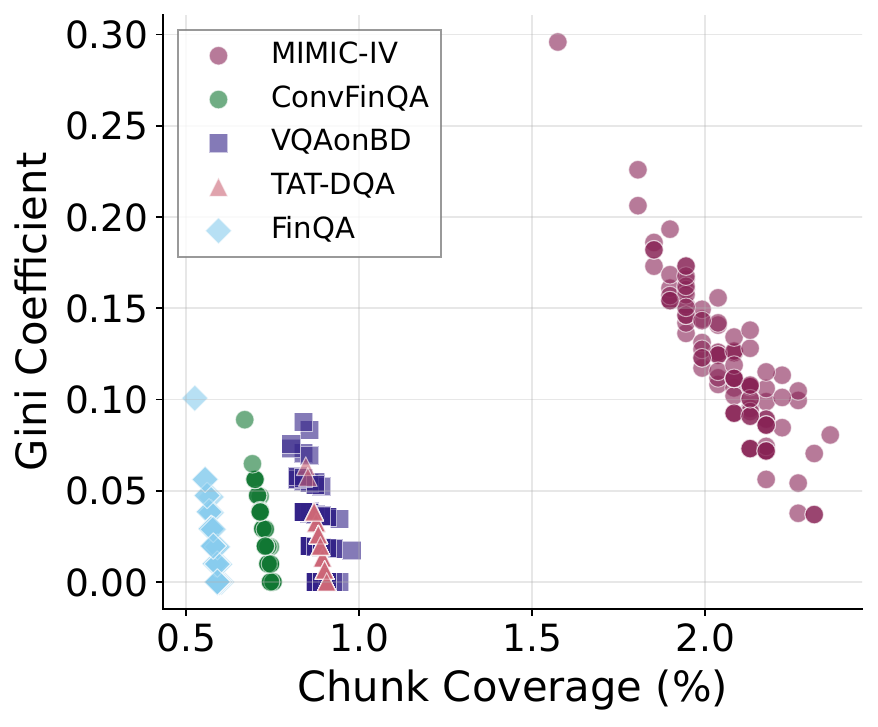}
        \caption{Gini Coefficient} \label{fig:rq1-scatter-gini}
    \end{subfigure}
    \hfill
    \begin{subfigure}[b]{0.32\textwidth}
        \centering
        \includegraphics[width=\linewidth]{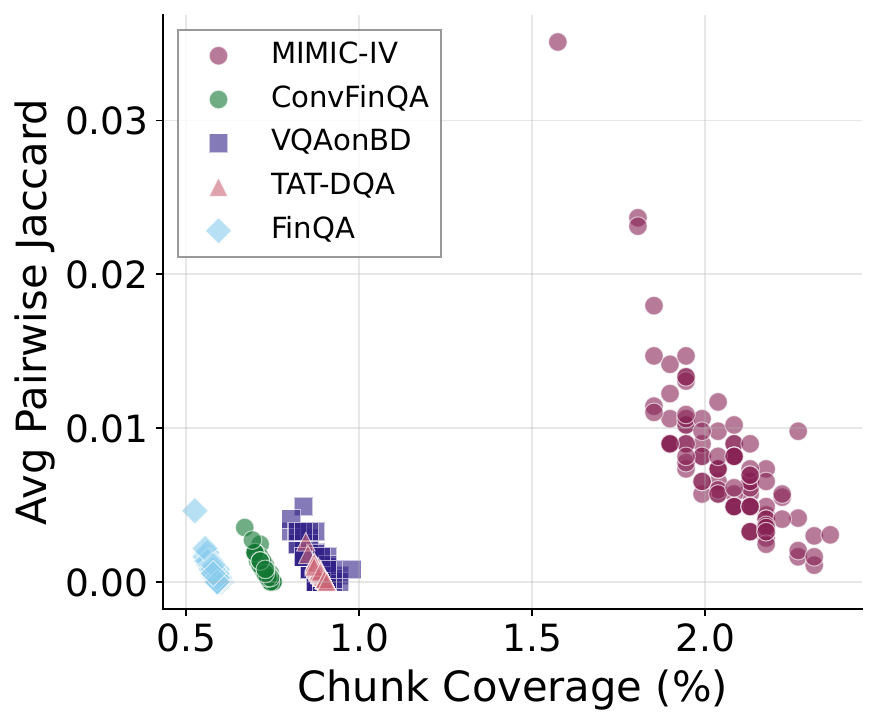}
        \caption{Jaccard Similarity} \label{fig:rq1-scatter-jaccard}
    \end{subfigure}

    \caption{Relationship between Chunk Coverage and diversity metrics for randomly sampled test suites ($n=50$). \textit{Higher} Chunk Coverage corresponds to \textit{higher} entropy (a), \textit{lower} Gini coefficient (b), and \textit{lower} Jaccard similarity (c).}
    \label{fig:rq1-scatter}
\end{figure}

\subsection{Chunk Coverage and Retrieval Diversity (RQ1)}
\label{sec:rq1}

RQ1 evaluates whether CC captures meaningful diversity in retrieval behaviour. To this end, we examine the relationship between CC and three established diversity measures: normalised entropy of chunk retrieval frequencies, the Gini coefficient measuring concentration of access, and the average pairwise Jaccard similarity between retrieved chunk sets. For each dataset, we generated 100 random test suites of sizes $n \in \{30, 50, 100\}$ using uniform random sampling and computed CC together with the corresponding diversity measures.

Figure~\ref{fig:rq1-scatter} visualises the relationship between CC and each diversity measure for test suites of size $n=50$. Each point represents one randomly sampled test suite. Across all datasets, higher CC values are associated with higher normalised entropy and lower Gini coefficient and Jaccard similarity. The absence of discontinuities or threshold effects in the plots indicates that CC varies smoothly with the diversity measures, suggesting that it captures gradual changes in retrieval behaviour rather than isolated extremes.
Within each dataset, it is visually possible to recognise a linear relationship between CC and each diversity metric, with a positive slope for entropy (Figure~\ref{fig:rq1-scatter-entropy}) and a negative slope for the other two metrics (Figures~\ref{fig:rq1-scatter-gini}, ~\ref{fig:rq1-scatter-jaccard}). \new{This confirms a positive correlation with entropy and negative correlations with the Gini coefficient and Jaccard similarity.}

\begin{table}[t]
\centering
\caption{Pearson correlations between Chunk Coverage and diversity metrics. All correlations are statistically significant and in the expected direction. Significance: $^{*}p<0.05$, $^{**}p<0.01$, $^{***}p<0.001$.}
\label{tab:rq1-correlations}
\small
\begin{tabular}{rlrrr}
\toprule
\textbf{Size ($n$)} & \textbf{Dataset} & \textbf{CC vs. Entropy} & \textbf{CC vs. Gini} & \textbf{CC vs. Jaccard} \\
\midrule
30 & MIMIC-IV  & $0.828^{***}$ & $-0.902^{***}$ & $-0.836^{***}$ \\
 & ConvFinQA & $0.574^{***}$ & $-0.589^{***}$ & $-0.578^{***}$ \\
 & FinQA     & $0.317^{**}$  & $-0.441^{***}$ & $-0.313^{**}$ \\
 & TAT-DQA   & $0.223^{*}$   & $-0.221^{*}$   & $-0.214^{*}$ \\
   & VQAonBD   & $0.758^{***}$ & $-0.767^{***}$ & $-0.756^{***}$ \\

\midrule
50 & MIMIC-IV  & $0.800^{***}$ & $-0.910^{***}$ & $-0.817^{***}$ \\
 & ConvFinQA & $0.956^{***}$ & $-0.982^{***}$ & $-0.956^{***}$ \\
 & FinQA     & $0.942^{***}$ & $-0.950^{***}$ & $-0.930^{***}$ \\
 & TAT-DQA   & $0.986^{***}$ & $-1.000^{***}$ & $-0.976^{***}$ \\
   & VQAonBD   & $0.582^{***}$ & $-0.561^{***}$ & $-0.632^{***}$ \\

\midrule
100& MIMIC-IV  & $0.763^{***}$ & $-0.907^{***}$ & $-0.760^{***}$ \\
& ConvFinQA & $0.856^{***}$ & $-0.874^{***}$ & $-0.858^{***}$ \\
& FinQA     & $0.861^{***}$ & $-0.911^{***}$ & $-0.890^{***}$ \\
& TAT-DQA   & $0.974^{***}$ & $-1.000^{***}$ & $-0.966^{***}$ \\
   & VQAonBD   & $0.692^{***}$ & $-0.693^{***}$ & $-0.765^{***}$ \\

\bottomrule
\end{tabular}
\end{table}

Table~\ref{tab:rq1-correlations} quantifies these relationships using Pearson correlation coefficients. Across all datasets and suite sizes, CC correlates positively with normalised entropy and negatively with both the Gini coefficient and average Jaccard similarity. All 45 correlations are statistically significant ($p < 0.05$) and occur in the expected directions. For test suites of size 50 or larger, the correlations are consistently strong. Averaged across datasets, CC exhibits a positive correlation with entropy ($r = 0.853$) and strong negative correlations with both the Gini coefficient ($r = -0.881$) and Jaccard similarity ($r = -0.862$). These results indicate that CC closely tracks established diversity measures when sufficient variation in retrieval behaviour is present.

For smaller suites ($n=30$), correlations are weaker for some datasets, most notably TAT-DQA and FinQA. This effect reflects the limited variance in coverage achievable with small test suites: when only a small number of queries is sampled from a large chunk space, CC values are inherently constrained. Even in this setting, however, all correlations remain statistically significant and directionally consistent.

The statistical results and their visual counterparts show that CC behaves as expected with respect to established diversity measures across datasets and suite sizes. This alignment provides empirical support that CC captures retrieval diversity in a meaningful and interpretable manner, motivating its use in the subsequent experiments.

\begin{tcolorbox}[boxrule=0pt,frame hidden,sharp corners,enhanced,
borderline north={1pt}{0pt}{black},
borderline south={1pt}{0pt}{black},
boxsep=2pt,left=2pt,right=2pt,top=2.5pt,bottom=2pt]
\textbf{Answer to RQ1:}
Chunk Coverage aligns closely with established diversity measures across datasets and suite sizes. Both statistical correlations and visual analyses show that higher coverage corresponds to more uniform, less concentrated, and less overlapping retrieval behaviour, supporting the use of Chunk Coverage as a meaningful proxy for retrieval diversity.
\end{tcolorbox}

\subsection{Guiding Test Generation/Selection with Chunk Coverage (RQ2)}
\label{sec:rq2}

RQ2 examines whether CC can guide the construction of test suites that explore retrieval behaviour efficiently. We compare three strategies: a coverage-guided strategy that greedily selects queries maximising incremental coverage, an overlap-biased strategy that deliberately favours redundant retrieval behaviour by prioritising queries with high chunk overlap, and a random strategy that samples uniformly from the query pool.

\begin{figure}[t]
\centering
\begin{minipage}{1.0\linewidth}
\centering

\begin{subfigure}[b]{0.32\linewidth}
    \centering
    \includegraphics[width=\linewidth]{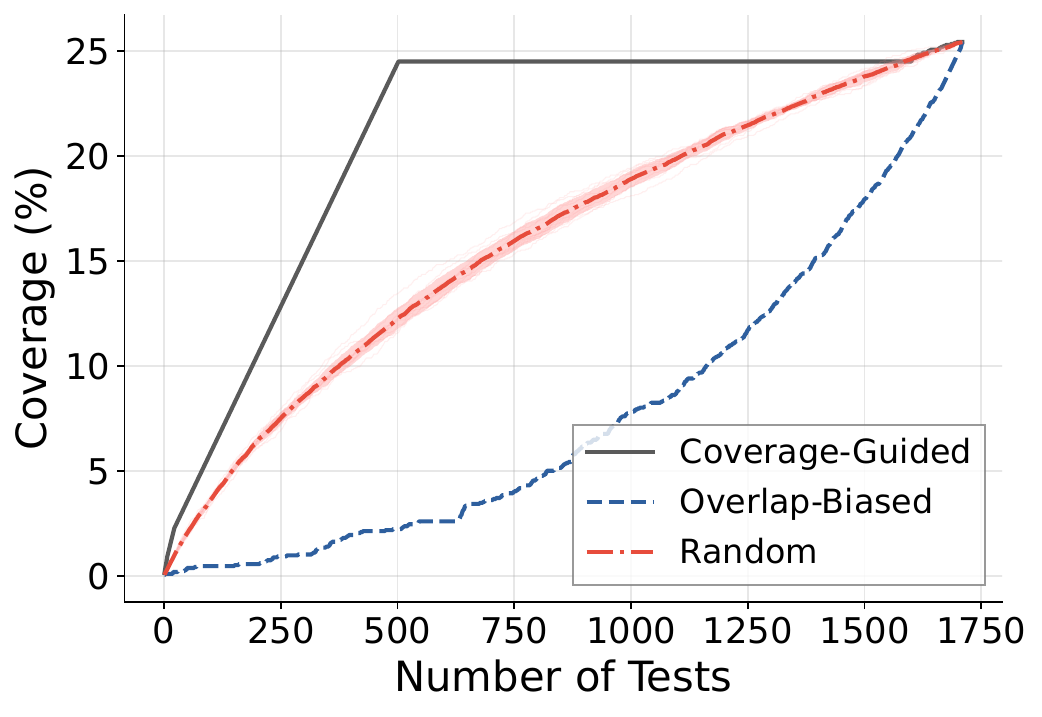}
    \caption{MIMIC-IV}
\end{subfigure}
\hfill
\begin{subfigure}[b]{0.32\linewidth}
    \centering
    \includegraphics[width=\linewidth]{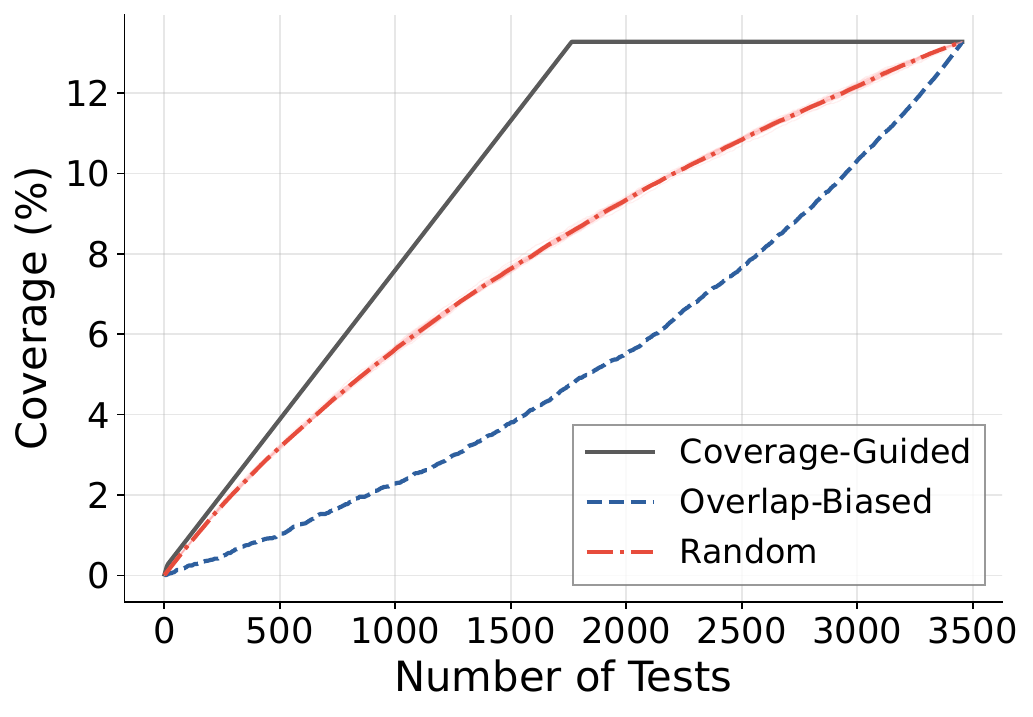}
    \caption{ConvFinQA}
\end{subfigure}
\hfill
\begin{subfigure}[b]{0.32\linewidth}
    \centering
    \includegraphics[width=\linewidth]{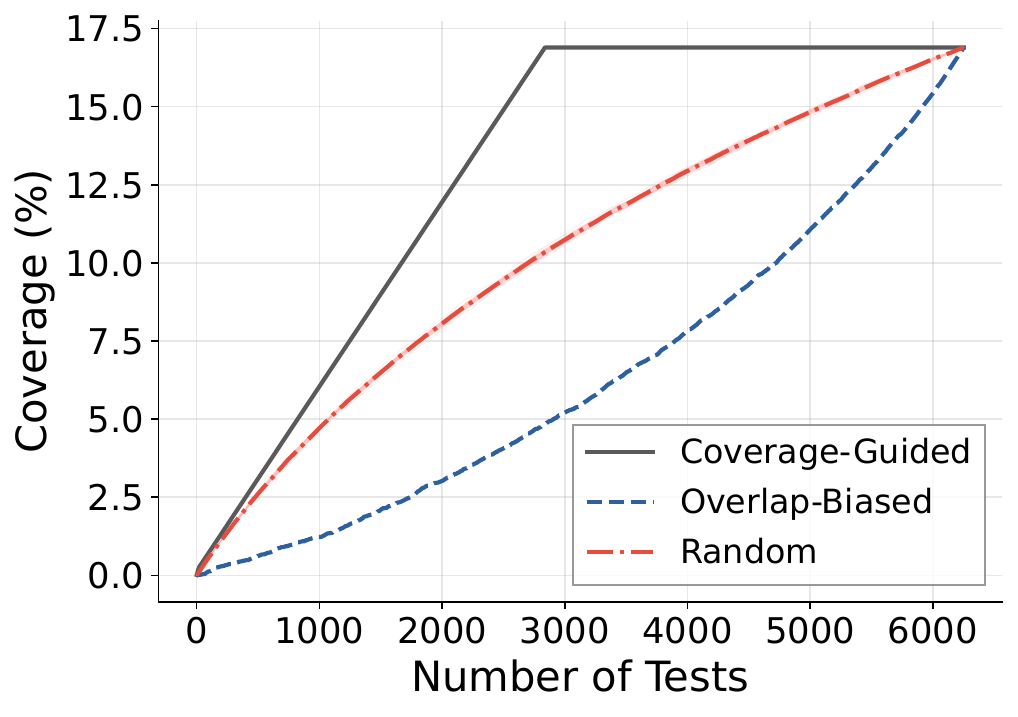}
    \caption{FinQA}
\end{subfigure}

\vspace{0.5em}

\begin{subfigure}[b]{0.32\linewidth}
    \centering
    \includegraphics[width=\linewidth]{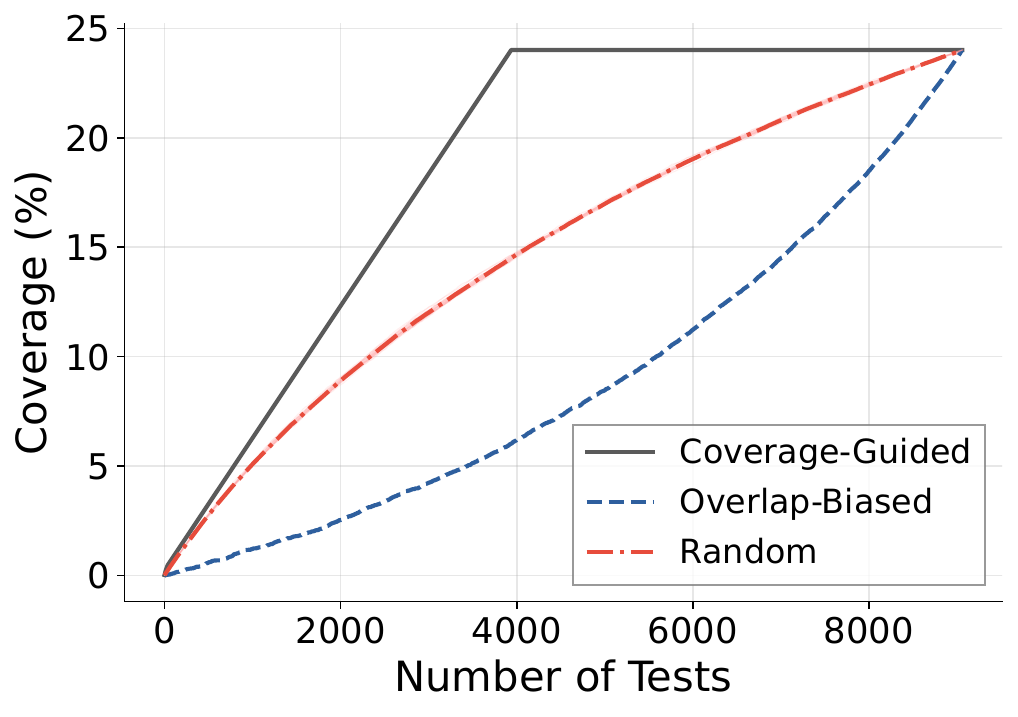}
    \caption{TAT-DQA}
\end{subfigure}
\hspace{0.02\linewidth} 
\begin{subfigure}[b]{0.32\linewidth}
    \centering
    \includegraphics[width=\linewidth]{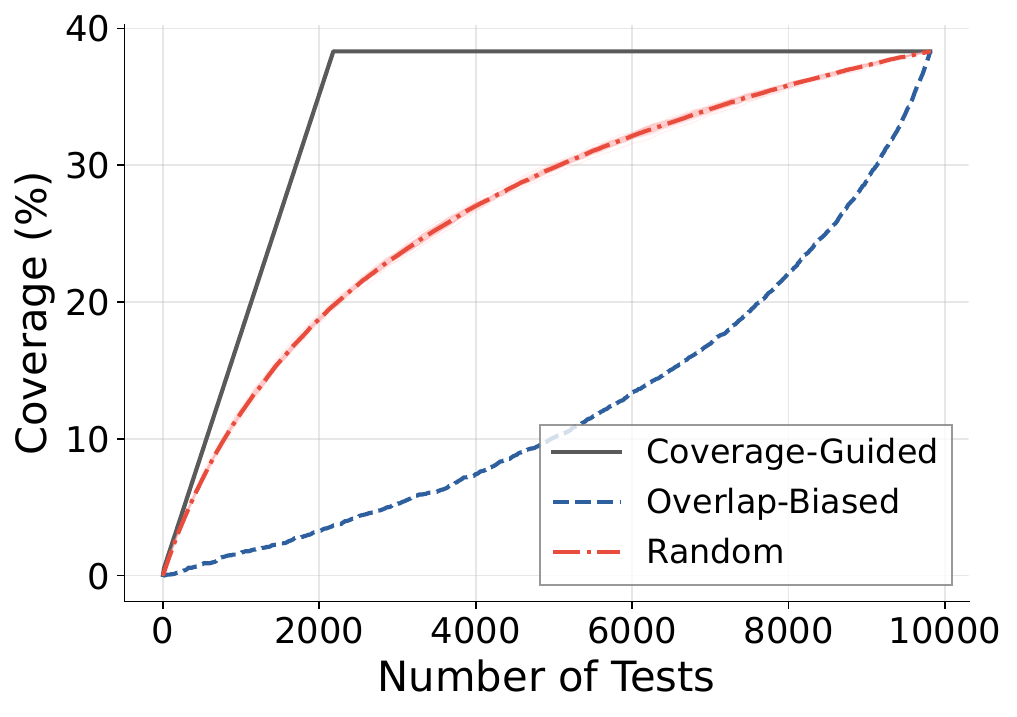}
    \caption{VQAonBD}
\end{subfigure}
\end{minipage}
\caption{Coverage growth under different test selection strategies. Coverage-guided selection consistently achieves faster coverage growth than random and overlap-biased selection.}
\label{fig:rq2-coverage}
\end{figure}

Figure~\ref{fig:rq2-coverage} shows coverage growth as a function of the number of executed tests. Across all datasets, coverage-guided selection consistently achieves faster coverage growth than both baselines, reaching a larger fraction of the attainable coverage at any given test budget. In contrast, the overlap-biased strategy exhibits slow growth, reflecting its tendency to repeatedly exercise the same retrieval behaviour.
A notable case is MIMIC-IV, where coverage-guided selection exhibits a prolonged stagnation (plateau) after approximately 500 executed tests, followed by a renewed increase in coverage after around 1,600 tests. This behaviour indicates that, during this interval, LLM-assisted test generation guided by uncovered chunks fails to elicit the retrieval of additional regions of the corpus. Only later do generated queries succeed in exercising previously unretrieved chunks, allowing coverage to increase again toward its maximum. We argue that such a plateau is plausible in realistic settings, where certain retrieval regions may be inherently difficult to reach or inaccessible due to limitations of LLMs in query formulation. The plateaus observed on the T2-RAGBench datasets arise because, although the available query pools are constructed to cover each document page, our retrieval space is defined at a finer granularity. Since each page is segmented into multiple chunks, page-level query coverage does not guarantee full chunk-level coverage, leaving some chunks inherently unreachable by the existing query pool.

Table~\ref{tab:rq2-speedup} reports the number of tests required by each strategy to reach 50\% of the maximum attainable CC. Across all datasets, the coverage-guided strategy consistently reaches this target with substantially fewer tests than both baselines. Relative to the random strategy, the coverage-guided strategy achieves a speedup between 1.4$\times$ and 2.1$\times$, and between 2.6$\times$ and 6.9$\times$ relative to the overlap-biased strategy. Averaged across datasets, this corresponds to a speedup of 1.7$\times$ over the random strategy and 4.2$\times$ over the deliberately redundant strategy. The effect is particularly pronounced for VQAonBD, where the relatively small chunk space amplifies the benefit of informed query strategy. The MIMIC-IV dataset exhibits similar behaviour, with coverage-guided strategy reaching 50\% coverage more than twice as fast as random strategy and over five times faster than overlap-biased strategy.

\begin{table}[t]
\centering
\caption{Number of tests required to reach 50\% of the maximum attainable coverage. Speedup reports the efficiency of coverage-guided selection relative to overlap-biased (OB) and random (R) strategies.}

\label{tab:rq2-speedup}
\small
\begin{tabular}{lrrrrr}
\toprule
\textbf{Dataset} & \textbf{Coverage-guided} & \textbf{Overlap-biased} & \textbf{Random} & \textbf{Speedup (OB)} & \textbf{Speedup (R)} \\
\midrule
MIMIC-IV & 248 & 1,301 & 530 & 5.2$\times$ & 2.1$\times$ \\
ConvFinQA & 871 & 2,259 & 1,226 & 2.6$\times$ & 1.4$\times$ \\
FinQA & 1,406 & 4,224 & 2,115 & 3.0$\times$ & 1.5$\times$ \\
TAT-DQA & 1,948 & 6,238 & 2,925 & 3.2$\times$ & 1.5$\times$ \\
VQAonBD & 1,080 & 7,468 & 2,096 & 6.9$\times$ & 1.9$\times$ \\
\midrule
Average & -- & -- & -- & 4.2$\times$ & 1.7$\times$ \\
\bottomrule
\end{tabular}
\end{table}

These results demonstrate that CC provides effective guidance for test generation. The consistently poor performance of the overlap-biased strategy confirms that CC penalises redundant retrieval behaviour, while the systematic advantage over the random strategy shows that CC enables principled prioritisation that accelerates exploration of the retrieval space.

\begin{tcolorbox}[boxrule=0pt,frame hidden,sharp corners,enhanced,
borderline north={1pt}{0pt}{black},
borderline south={1pt}{0pt}{black},
boxsep=2pt,left=2pt,right=2pt,top=2.5pt,bottom=2pt]
\textbf{Answer to RQ2:}
Chunk Coverage enables systematic and efficient test generation. Coverage-guided selection reaches 50\% of the attainable coverage 1.7$\times$ faster than random selection and 4.2$\times$ faster than deliberately redundant selection on average, while consistently achieving higher coverage at comparable test budgets.
\end{tcolorbox}

\subsection{Relation to Testing Effectiveness (RQ3)}
\label{sec:rq3}

RQ3 investigates whether coverage-guided strategy improves fault detection effectiveness, that is, whether it enables faults to be discovered earlier during testing. As introduced in Section~\ref{sec:failures_and_faults}, we distinguish between failures, which are observable violations of intended RAG behaviour identified through extreme evaluation outcomes, and faults, which are distinct root causes inferred by grouping failures that exhibit highly overlapping retrieval results. To quantify fault detection effectiveness, we adopt APFD, a standard metric from regression testing that captures how early faults are detected in a test execution order. Higher APFD values indicate that faults tend to be revealed earlier. \new{Unlike the 50\%-coverage prefixes in Table~\ref{tab:rq2-speedup}, Table~\ref{tab:rq3-apfd} computes APFD over the same complete query pool for all strategies, so differences reflect only prioritisation.} \new{APFD is appropriate here because every strategy executes the complete query set, every query has the same cost (one retrieval and one LLM generation call under the same configuration), and our failure definition assigns binary severity.} Context precision is computed differently across datasets. As MIMIC-IV does not provide ground-truth relevance annotations for retrieved chunks, context precision is estimated using an LLM-as-a-judge without reference annotations, reported as \textit{Context Precision (w/o GT)}. In contrast, T2-RAGBench provides page-level relevance annotations. In this setting, a retrieved chunk is considered relevant if it originates from a ground-truth annotated page, enabling context precision to be evaluated against reference information.

\begin{table*}[t]
\centering
\caption{Fault detection effectiveness (APFD) by dataset and RAG evaluation metric. OB, R, and CC denote overlap-biased, random, and coverage-guided strategies, respectively. Best values are highlighted in \textbf{bold}.}
\label{tab:rq3-apfd}
\small
\begin{tabular}{llrrccc}
\toprule
\textbf{Dataset} & \textbf{RAG Eval. Metric} & \textbf{\# Failures} & \textbf{\# Faults} 
& \textbf{OB} & \textbf{R} & \textbf{CC} \\
\midrule
\multirow{5}{*}{MIMIC-IV}
 & Faithfulness                       & 56  & 38  & 0.458 & 0.562 & \textbf{0.566} \\
 & Response Relevancy                 & 495 & 199 & 0.362 & 0.574 & \textbf{0.700} \\
 & Context Relevance                  & 377 & 103 & 0.381 & 0.584 & \textbf{0.697} \\
 & Response Groundedness              & 228 & 41  & 0.523 & \textbf{0.615} & 0.597 \\
 & Context Precision (w/o GT)         & 314 & 116 & 0.383 & 0.563 & \textbf{0.688} \\
\midrule
\multirow{6}{*}{ConvFinQA}
 & Faithfulness                       & 834 & 587 & 0.456 & 0.547 & \textbf{0.617} \\
 & Context Precision                  & 2{,}302 & 1{,}228 & 0.424 & 0.583 & \textbf{0.704} \\
 & Response Relevancy                 & 1{,}623 & 952 & 0.431 & 0.570 & \textbf{0.669} \\
 & Context Relevance                  & 1{,}466 & 857 & 0.438 & 0.574 & \textbf{0.657} \\
 & Response Groundedness              & 1{,}684 & 973 & 0.435 & 0.573 & \textbf{0.679} \\
 & Context Precision (w/o GT)         & 1{,}686 & 989 & 0.430 & 0.571 & \textbf{0.673} \\
\midrule
\multirow{6}{*}{FinQA}
 & Faithfulness                       & 1{,}736 & 1{,}127 & 0.426 & 0.558 & \textbf{0.627} \\
 & Context Precision                  & 4{,}195 & 1{,}986 & 0.406 & 0.597 & \textbf{0.711} \\
 & Response Relevancy                 & 2{,}140 & 1{,}259 & 0.418 & 0.567 & \textbf{0.652} \\
 & Context Relevance                  & 2{,}763 & 1{,}500 & 0.415 & 0.578 & \textbf{0.675} \\
 & Response Groundedness              & 2{,}808 & 1{,}570 & 0.413 & 0.574 & \textbf{0.670} \\
 & Context Precision (w/o GT)         & 2{,}567 & 1{,}446 & 0.413 & 0.569 & \textbf{0.664} \\
\midrule
\multirow{6}{*}{TAT-DQA}
 & Faithfulness                       & 1{,}897 & 1{,}236 & 0.441 & 0.558 & \textbf{0.616} \\
 & Context Precision                  & 4{,}651 & 2{,}220 & 0.388 & 0.593 & \textbf{0.724} \\
 & Response Relevancy                 & 2{,}849 & 1{,}684 & 0.412 & 0.566 & \textbf{0.660} \\
 & Context Relevance                  & 2{,}876 & 1{,}622 & 0.419 & 0.571 & \textbf{0.654} \\
 & Response Groundedness              & 3{,}405 & 1{,}912 & 0.416 & 0.572 & \textbf{0.658} \\
 & Context Precision (w/o GT)         & 2{,}910 & 1{,}682 & 0.417 & 0.569 & \textbf{0.659} \\
\midrule
\multirow{6}{*}{VQAonBD}
 & Faithfulness                       & 1{,}749 & 782 & 0.421 & 0.597 & \textbf{0.720} \\
 & Context Precision                  & 7{,}141 & 1{,}778 & 0.351 & 0.669 & \textbf{0.839} \\
 & Response Relevancy                 & 2{,}816 & 1{,}007 & 0.394 & 0.622 & \textbf{0.761} \\
 & Context Relevance                  & 3{,}291 & 1{,}016 & 0.395 & 0.641 & \textbf{0.780} \\
 & Response Groundedness              & 3{,}912 & 1{,}329 & 0.378 & 0.631 & \textbf{0.779} \\
 & Context Precision (w/o GT)         & 2{,}774 & 976 & 0.402 & 0.624 & \textbf{0.765} \\
\bottomrule
\end{tabular}
\end{table*}

Table~\ref{tab:rq3-apfd} reports fault detection effectiveness across datasets and RAG evaluation metrics. For each dataset and metric, we report the total number of observed failures, the number of distinct faults obtained by clustering failures at the retrieval level, and the resulting APFD for coverage-guided (CC), random (R), and overlap-biased (OB) strategies. Across nearly all datasets and evaluation metrics, the coverage-guided strategy achieves the highest APFD, indicating that faults are detected earlier than with either baseline. This trend is particularly consistent in T2-RAGBench. For example, on VQAonBD, coverage-guided strategy achieves APFD values between 0.720 and 0.839 across metrics, compared to 0.597--0.669 for the random strategy. Similar improvements are observed on ConvFinQA, FinQA, and TAT-DQA, where the coverage-guided strategy consistently outperforms random and overlap-biased strategies across all reported metrics.

For MIMIC-IV, one exception occurs for response groundedness, where the random strategy slightly outperforms the coverage-guided strategy when test queries are generated using GPT-4o-mini. Since test queries for MIMIC-IV are obtained through LLM-assisted generation rather than from an existing, curated query set, the effectiveness of coverage-guided selection is bounded by the ability of the generation model to produce queries that meaningfully target distinct retrieval behaviour. To assess whether this effect is specific to the choice of generation model, we repeated the MIMIC-IV experiment using a stronger generator (GPT-5-mini) for test generation. In this setting, coverage-guided selection consistently outperforms random selection across all evaluation metrics, including response groundedness. This additional result suggests that the earlier exception can be attributable to limitations of the test generation model rather than to CC itself. We include the full results of this supplementary experiment in the online artefact (see Section~\ref{sec:data}).

% In contrast to T2-RAGBench, where queries are accompanied by ground-truth answers and relevance annotations, we suspect that LLM-generated queries may provide weaker guidance for exercising the retriever. Under such conditions, the benefits of CC can be attenuated, allowing random selection to perform comparably or, in isolated cases, slightly better. Exploring the impact of alternative query generation models is left to future work.

Overall, these results show that CC is closely aligned with testing effectiveness. By prioritising queries that exercise previously unvisited retrieval behaviour, the coverage-guided strategy advances fault discovery to earlier stages of testing. In contrast, strategies that favour redundant retrieval behaviour delay exposure of new failure modes. \new{Because RQ3 includes LLM-output failures (Faithfulness, Response Relevancy, and Response Groundedness), the APFD gains indirectly link broader retrieval coverage to diverse output failures, but CC neither measures nor guarantees output correctness. A causal analysis remains future work.}

\begin{tcolorbox}[boxrule=0pt,frame hidden,sharp corners,enhanced,
borderline north={1pt}{0pt}{black},
borderline south={1pt}{0pt}{black},
boxsep=2pt,left=2pt,right=2pt,top=2.5pt,bottom=2pt]
\textbf{Answer to RQ3:}
Chunk Coverage is positively associated with testing effectiveness. Across datasets and evaluation metrics, coverage-guided selection detects distinct faults earlier than both random and overlap-biased strategies, improving fault detection effectiveness (APFD) by approximately 10\% to 25\%\footnotemark{} relative to random selection.
These results indicate that prioritising retrieval diversity accelerates fault discovery without requiring explicit test oracles.
\end{tcolorbox}

\footnotetext{Improvements are computed as relative APFD gains over the random baseline, i.e., $(\mathrm{APFD}_{\text{CC}} - \mathrm{APFD}_{\text{R}})/\mathrm{APFD}_{\text{R}}$. The minimum observed improvement ($\approx 10.4\%$) occurs for Faithfulness on TAT-DQA $(0.616-0.558)/0.558$, while the maximum observed improvement ($\approx 25.4\%$) occurs for Context Precision on VQAonBD $(0.839-0.669)/0.669$.}

% \input{sections/discussion}
% !TEX root = ../main.tex

\section{Related Work}
\label{sec:related_work}

\subsection{Test Adequacy and Coverage Criteria}

Test adequacy criteria have long played a central role in software testing, providing a principled basis for assessing whether a test suite sufficiently exercises a system under test~\cite{ammann2017introduction}. In traditional software, structural coverage metrics such as statement, branch, and path coverage approximate the extent to which program behaviour has been exercised, independently of output correctness~\cite{zhu1997software}. These metrics do not guarantee the absence of faults, but they provide actionable feedback for systematically improving test suites. In contrast, mutation testing assesses adequacy by measuring a test suite’s ability to detect injected faults through behavioural differences between the original program and its mutants, even when an explicit oracle for the original output is unavailable~\cite{papadakis2019mutation}.

As Machine Learning (ML) systems gained prominence, several works extended the notion of coverage beyond control-flow structures to model-internal representations. DeepXplore~\cite{pei2017deepxplore} and DeepGauge~\cite{ma2018deepgauge} propose adequacy criteria based on neuron activations, aiming to quantify how thoroughly a deep neural network is exercised by a test set. Surprise Adequacy~\cite{kim2019guiding,kim2023evaluating} further characterises adequacy in terms of the novelty of test inputs with respect to the model’s training distribution, capturing how far test executions deviate from previously observed behaviours. Collectively, these approaches move beyond output accuracy as the sole evaluation criterion and instead emphasise behavioural exploration, in line with the role of coverage in traditional software testing.

\new{CC follows the same structural-adequacy principle but covers externally retrieved chunks rather than program structures or model-internal activations.}

\subsection{Testing and Evaluation of LLMs}

Although CC does not rely on LLMs and exclusively targets the retrieval component of RAG systems, LLMs remain a central element in most practical RAG deployments. We therefore briefly relate our work to prior research on testing and evaluating LLMs in order to situate CC within the broader landscape of testing methodologies for language-based systems. Recent work on LLM evaluation has examined a wide range of failure modes, including robustness through behavioural and adversarial testing~\cite{ribeiro2020beyond,nie2020adversarial,kiela2021dynabench}, as well as bias and fairness using targeted benchmarks and audit-style evaluations~\cite{gallegos2024bias}. Other studies have focused on hallucination and factuality, particularly in knowledge-intensive and open-domain settings~\cite{lin2022truthfulqa,ji2023survey}. These approaches typically assess models using per-input performance metrics or behavioural probes, which are often aggregated over curated benchmark datasets. While effective at revealing specific weaknesses, such evaluations do not provide explicit notions of test adequacy at the level of a test suite.  Coverage-oriented perspectives have begun to emerge for language models. Frameworks such as CheckList~\cite{ribeiro2020beyond} organise tests around linguistic capabilities, and adaptive testing approaches~\cite{ribeiro2022adaptive} iteratively generate tests based on observed failures. These works promote a coverage-oriented mindset but remain centred on the language model itself and do not account for systems whose behaviour is shaped by external retrieval components.

In contrast to these lines of work, CC targets the retrieval behaviour of RAG systems. Such behaviour is neither captured by traditional code coverage nor by model-internal coverage metrics. Rather than characterising how a model processes inputs internally, CC characterises how a test suite exercises the externally induced behaviour space defined by the document corpus, the embedding model, and the retrieval configuration.

\section{Threats to Validity}
\label{sec:threats}

\paragraph{Construct validity} CC abstracts test adequacy in terms of retrieval behaviour and does not assess output correctness. While this mirrors structural coverage criteria in traditional testing, it may over-approximate adequacy when some retrieved chunks are irrelevant to the task. We mitigate this by focusing on retrieval-dependent RAG scenarios and by interpreting CC as a testing guidance signal rather than a correctness measure. 

\paragraph{Internal validity} LLM inference and random test selection introduce non-determinism. We mitigate this by repeating non-deterministic experiments and reporting averaged results. For MIMIC-IV, test queries are generated using an auxiliary LLM; limitations in query quality may reduce the observable benefit of CC. To control for this, all evaluation strategies operate on the same fixed query pool once generated.

\paragraph{External validity} Our study focuses on two high-stakes, retrieval-dependent domains, and the results may not generalise to open-domain RAG settings where retrieval is optional. Moreover, the observed results depend on specific RAG configurations, such as chunking and embedding choices. However, CC is not intended for comparing different RAG system configurations, just as traditional coverage criteria are not used to compare different program implementations, but rather to guide test selection for a given system under test.

% !TEX root = ../main.tex

\section{Conclusion}
\label{sec:conclusion}

We presented \textit{Chunk Coverage (CC)} as a test adequacy criterion for the retrieval component of RAG systems. CC exposes retrieval behaviour at the level of a test suite, independently of per-query correctness, and provides a practical signal for guiding test generation/selection without requiring test oracles. Across clinical and financial RAG scenarios, CC-guided testing systematically explores a larger portion of the retrieval space and detects distinct retrieval faults earlier than random or redundancy-biased strategies. These results show that retrieval diversity, as captured by CC, is directly linked to testing effectiveness. By complementing existing per-query RAG evaluation metrics with a suite-level perspective, CC offers a principled foundation for systematic testing and validation of retrieval-centric LLM systems prior to deployment.

\section*{Data Availability}
\label{sec:data}
The implementation code for chunk coverage and the data required to reproduce the experiments are available at: \url{https://github.com/dbr7/issta26-chunk-coverage-artifact}.

\bibliographystyle{ACM-Reference-Format}
\bibliography{bibliography}

\end{document}